\documentclass[%
 reprint,
superscriptaddress,
 amsmath,amssymb,
 aps,
]{revtex4-2}

\usepackage{graphicx}% Include figure files
\usepackage{dcolumn}% Align table columns on decimal point
\usepackage{bm}% bold math
\usepackage[mathlines]{lineno}% Enable numbering of text and display math
\begin{document}

% \preprint{APS}  %/123-QED

\title{How close are the classical two-body potentials to ab initio calculations? Insights from linear machine learning based force matching}% Force line breaks with \\
% % \thanks{A footnote to the article title}%

\author{Zheng Yu}
\affiliation{Department of Chemistry, University of Illinois at Urbana-Champaign, Urbana, 61801, USA}
%  \altaffiliation[Also at ]{Physics Department, XYZ University.}%Lines break automatically or can be forced with \\
\author{Ajay Annamareddy}
\affiliation{Department of Materials Science and Engineering, University of Wisconsin-Madison, Madison, 53706, USA}
\author{Dane Morgan}
\affiliation{Department of Materials Science and Engineering, University of Wisconsin-Madison, Madison, 53706, USA}
\author{Bu Wang}%
 \email{bu.wang@wisc.edu}
 \affiliation{Department of Materials Science and Engineering, University of Wisconsin-Madison, Madison, 53706, USA}
 \affiliation{%
 Department of Civil and Environmental Engineering, University of Wisconsin-Madison, Madison, 53706, USA
}%


\begin{abstract}

In this work, we propose a linear machine learning force matching approach that can directly extract pair atomic interactions from ab initio calculations in amorphous structures. The local feature representation is specifically chosen to make the linear weights a force field as a force/potential function of the atom pair distance. Consequently, this set of functions is the closest representation of the ab initio forces given the two-body approximation and finite scanning in the configurational space. We validate this approach in amorphous silica. Potentials in the new force field (consisting of tabulated Si-Si, Si-O, and O-O potentials) are significantly softer than existing potentials that are commonly used for silica, even though all of them produce the tetrahedral network structure and roughly similar glass properties. This suggests that those commonly used classical force fields do not offer fundamentally accurate representations of the atomic interaction in silica. The new force field furthermore produces a lower glass transition temperature ($T_g\sim$1800 K) and a positive liquid thermal expansion coefficient, suggesting the extraordinarily high $T_g$ and negative liquid thermal expansion of simulated silica could be artifacts of previously developed classical potentials. Overall, the proposed approach provides a fundamental yet intuitive way to evaluate two-body potentials against ab initio calculations, thereby offering an efficient way to guide the development of classical force fields. 

\end{abstract}

\maketitle

% \tableofcontents

\section{\label{sec:intro}Introduction}

Computer modeling of atoms and molecules is an indispensable tool today for probing atomic-level physics, understanding materials behaviors, deciphering chemical reactions, and examining biological processes.\cite{jorgensenComparisonSimplePotential1983,kollmanCalculatingStructuresFree2000} At the heart of these endeavors lies the intricate task of accurately and efficiently characterizing interactions among atoms or molecules.\cite{rapaport2004art} Ab initio methods, such as the wavefunction methods and the density functional theory (DFT), offer rigorous calculations of these interactions based on the quantum mechanics of electrons coupled with the Born-Oppenheimer approximation for nuclei treatment, generally delivering top-tier accuracy.\cite{foulkesQuantumMonteCarlo2001,jonesDensityFunctionalTheory2015} However, their demanding computational nature restricts their applicability, often confining simulations to a few hundred atoms over relatively short timescales (tens to hundreds of picosecond) with standard hardware. In contrast, classical molecular dynamics and Monte Carlo simulations, which employ empirical atomic interactions typically expressed as functions of atomic distances or angles, offer substantial computational efficiency, facilitating simulations of millions of atoms and the millisecond timescale. However, this efficiency often comes at the expense of accuracy and generalizability. 

Navigating the tradeoff between computational accuracy and  efficiency remains a paramount challenge in atomic-scale modeling.\cite{friederichMachinelearnedPotentialsNextgeneration2021} The emergence of machine-learning force fields holds promise, endeavoring to marry the accuracy of ab initio methods with affordable computational cost, e.g., artificial (and graph) neural network potentials, Gaussian approximation potentials, moment tensor potentials, and atomic cluster expansion potentials \cite{behlerGeneralizedNeuralNetworkRepresentation2007,bartokGaussianApproximationPotentials2010,thompsonSpectralNeighborAnalysis2015,shapeevMomentTensorPotentials2016,attarianThermophysicalPropertiesFLiBe2022,drautzAtomicClusterExpansion2019}. Yet, while they introduce innovations, significant issues remain: 1) they are typically slower than simple pair potentials by around 1-2 orders of magnitude; 2) they provide less explicit and easy-to-access physical insights about the many-body interactions; 3) they risk more failure away from the training region.\cite{erhardMachinelearnedInteratomicPotential2022,zuoPerformanceCostAssessment2020} While solutions to those issues are being developed, classical interatomic potentials remain the dominant choice across the molecular simulation communities, especially for large and complex systems such as biological systems and disordered materials.

Classical potentials utilize pre-defined function forms, such as Lennard-Jones, to describe the atomic interactions. Parameters of the function form are optimized to align simulation results with observables sourced from either experimental data or ab initio simulations. Observables used most often include structural features (e.g., pair distribution functions for disordered systems and unit cell structure for crystals) and physical properties (e.g., density and mechanical properties). Because classical potentials are approximations of the true atomic interactions, such a potential fitting practice does not yield a unique solution. Combinations of wildly different parameters may give very similar fitting errors; they may all reproduce the included observables but can lead to very different behaviors in actual simulations. Most force fields for ionic systems also define Coulomb interactions using arbitrarily selected or ill-defined point charges, adding more to the arbitrariness. One example is silica (SiO$_2$), an archetypal disordered material. The well-known Beest-Kramer-van Santen (BKS) potential, probably the most used potential for silica, produces density and glass transition temperatures that are in poor agreement with experiments.\cite{yuStructuralSignaturesThermodynamic2021} Newer potentials like Sundarararaman-Huang-Ispas-Kob (SHIK) have been developed by adding more observables at various pressures.  \cite{vanbeestForceFieldsSilicas1990,sundararamanNewOptimizationScheme2018} There also exist several versions of widely used force fields for modified silicate glasses, each of which includes a different version of potential for SiO$_2$.\cite{pedoneNewSelfConsistentEmpirical2006,wangNewTransferableInteratomic2018,dengDevelopmentBoronOxide2019} There have been many studies trying to evaluate the accuracy of the those force fields, most of which are nonetheless limited to comparing the simulated properties/behaviors to experiments or ab inito simulations. There is a lack of fundamental insights into how accurate those empirical potentials are in describing atomic interactions. There also exist questions about the transferability of these classical potentials. For instance, how reasonable is it to use the same silica potentials in modified silicates?

In this work, we introduce a direct method to evaluate pair interactions from ab initio calculations by linear machine learning (ML) based force matching. The method is intuitive and fast, requiring only a small amount of ab initio data. It produces the closest approximation to the ab initio forces given the two-body approximation and finite scanning in the configurational space. When applied to silica, we found that commonly used classical force fields do not offer fundamentally accurate representations of atomic interactions. The force matching method can also generate classical pair potentials without pre-defined function forms and parameter optimization. The new potential generated for silica yield properties much closer to experimental observations, especially in the liquid region. Furthermore, we demonstrate how to apply this method to evaluate the transferability of silica potentials to sodium silicates and borosilicates. Overall, this approach provides explicit physical insights into the atomic interactions and can enable fast and automatic development of new classical pair potentials.

\section{Methods}
\subsection{Force matching regression method}\label{sec:method_ML}

In classical MD simulations, the potential ideally approximates the exact many-body atomic interaction seen in quantum mechanics over the configurational space, by achieving
\begin{equation}\label{eq:U}
    \min_{U_\mathrm{MD}} \int d\mathbf{R}^N |U_\mathrm{MD}(\mathbf{R}^N)-U_\mathrm{QM}(\mathbf{R}^N)|^2
\end{equation}
where $U_\mathrm{MD}$ and $U_\mathrm{QM}$ are the potential energies in classical MD and in quantum mechanics, respectively, and both are functions of $\mathbf{R}^N$, the coordinates of all $N$ atoms in the system. Since $\mathbf{F}_I = -\partial U(\mathbf{R}^N)/\partial \mathbf{R}_I$ where $\mathbf{F}_I$ is the force on atom $I$ by its surrounding atoms and $\mathbf{R}_I$ is atom $I$'s coordinate, we can recast Equation \ref{eq:U} into its equivalent form for forces\cite{ercolessiInteratomicPotentialsFirstPrinciples1994} 
\begin{equation}
    \min_{F_\mathrm{MD}} \int d\mathbf{R}^N \sum_I |\mathbf{F}_I^\mathrm{MD}(\mathbf{R}_I;\mathbf{R}^N)-\mathbf{F}_I^\mathrm{QM}(\mathbf{R}_I;\mathbf{R}^N)|^2
\end{equation}
Therefore, the potential development can be solved as a force matching problem.

The most elementary form of $\mathbf{F}_I^\mathrm{MD}$ is pair (2-body) forces, i.e., 
\begin{equation}
    \mathbf{F}_{I}^\mathrm{MD}(\mathbf{R}_{I};\mathbf{R}^N) \simeq \sum_{J\neq I} \mathbf{F}_{IJ}^\mathrm{MD}(\mathbf{R}_I,\mathbf{R}_J)= \sum_{J\neq I}\mathbf{F}_{IJ}^\mathrm{MD}(\mathbf{R}_{IJ})
\end{equation}
where $J$ denotes any other atom in the system (within a cutoff distance in practice) and $\mathbf{R}_{IJ}$ is the vector directed towards atom $I$ from atom $J$. This pair approximation can be adequate in many materials, even when bonds of covalent nature are present. One particular example is the main group oxide. The strong network-forming covalent bonds in these materials restrict the local degrees of freedom and enable effective decomposition of multi-body interactions as functions of pair distances. This promotes the popularity of pair potentials for classical MD simulations in these contexts.\cite{matsuiTransferableInteratomicPotential1994,pedoneNewSelfConsistentEmpirical2006} 

Within the pair approximation, forces are simplified to functions of pair distances without concern for orientations. Hence, the forces on individual atoms can be rewritten as 
\begin{equation}\label{eq:force}
    \mathbf{F}_I^\mathrm{MD}(\mathbf{R}_I;\mathbf{R}^N) \simeq \int_0^{r_\mathrm{cut}} dr f(r) \sum_{J\neq I} \delta(r-|\mathbf{R}_{IJ}|) \hat{u}(\mathbf{R}_{IJ})
\end{equation}
where $f(r)$ is the pair force field as a function of distance from zero to a cutoff distance $r_\mathrm{cut}$, $|\mathbf{R}_{IJ}|$ and $\hat{u}(\mathbf{R}_{IJ})$ are the length and the unit vector of $\mathbf{R}_{IJ}=\mathbf{R}_{I}-\mathbf{R}_{J}$, respectively, and $\delta$ is the Dirac delta function. For simplicity, Equation \ref{eq:force} considers only one atom species but it can be easily extended to multi-species systems with different $f_{ab}(r)$ pairs, where $a,b$ denotes atomic species. 

To make Equation \ref{eq:force} applicable in simulations, the integral can be discretized by summing over thin spherical shells between from 0 to $r_\mathrm{cut}$, that is
\begin{equation}\label{eq:vector}
    \mathbf{F}_I^\mathrm{MD}(\mathbf{R}_I;\mathbf{R}^N) \simeq \sum_{i (\mathrm{shell})} f(r_i) \sum_{J_i} \hat{u}(\mathbf{R}_{IJ_i})
\end{equation}
where $i$ denotes the $i$th shell, $r_i$ is the radius of the shell, and $J_i$ denotes an atom index within the $i$th shell. Projected onto a basis and considering all species, the above formula becomes 
\begin{equation}
\begin{aligned}\label{eq:core}
    F_{aI,x}^\mathrm{MD}(\mathbf{R}_{aI};\mathbf{R}^N) \simeq & \sum_{b} \sum_{i (\mathrm{shell})} f_{ab}(r_i) \sum_{bJ_i} \hat{u}(\mathbf{R}_{aI,bJ_i})\cdot \hat{x} \\ 
    = & \sum_{b} \sum_{i (\mathrm{shell})} f_{ab}(r_i) X_{aI}[b,i] \\
    = & \mathcal{F} X_{aI}
\end{aligned}
\end{equation}
where $\hat{x}$ is the unit vector along $x$ axis. 

As derived in Equation \ref{eq:core}, the force on atom $I$ holds an approximately linear relationship with the feature vector $X_{aI}$ concatenated by elements of $X_{aI}[b,i]=\sum_{bJ_i} \hat{u}(\mathbf{R}_{aI,bJ_i})\cdot \hat{x}$, which can be understood as an effective `number' of neighboring atoms in the corresponding shell depending on the orientations. Thus, the linear coefficient vector $\mathcal{F}$, concatenated by $f_{ab}(r_i)$, is a vector of effective forces for all species pairs as functions of pair distances, which is exactly the target force field for classical MD simulations but in a tabulated form. As the shells becomes thinner, the tabulated force field, consisting of a series of $f_{ab}(r_i)$, approaches the continuous $f_{ab}(r)$ in Equation \ref{eq:force}. In addition, this representation can also be used with the kernel tricks and other ML methods, but it will not be as straightforward as linear regression to extract the effective pair force fields from ab initio calculations. 

The coefficients (or weights) in the simple linear regression can be solved analytically as $(X^\top X)^{-1}XF$. However, the coefficients could be easily biased to the training data resulting in overfitting, especially as the dimension of $X$ is large. Therefore, to solve this force-matching problem, we use the Ridge regression ML model (with Tikhonov regularization).\cite{hoerlRidgeRegressionBiased1970} Instead of directly minimizing $\sum_{a,I,x,C} |F_{aI,x}^\mathrm{MD}-F_{aI,x}^\mathrm{QM}|^2$ in different configurations $C$, the model aims to minimize $\sum_{a,I,x,C} [F_{aI,x}^\mathrm{MD}(f_{ab}(r_i))-F_{aI,x}^\mathrm{QM}]^2 + \lambda \sum_{ab,i} f_{ab}(r_i)^2$, where $\lambda$ is the regularization parameter controlling the constraints (or degrees of freedom) in the force field space. Note that $\lambda$ is a hyperparameter that is determined by cross validation before training the final model. 

Once the ML model is trained, the weights of the model as a function of distance can be directly used as a classical force field, without further parametrization and optimization. When applying this force field to MD simulations, the ML model or any extra descriptor calculation is not needed. This is a major difference between this approach and ML potentials---the resulting force field from this approach has a format similar to classical potentials and therefore is as efficient in MD simulations. 

\subsection{Additional settings for the force field}\label{sec:potentialsetting}

Like other classical potentials, the potentials generated in this work have a long-range cutoff, beyond which no interactions are considered explicitly. The long-range cutoff $r_{cut}^{LR}$ is set to 8 {\AA} in this study for two reasons: (1) atomic interactions are found to be very weak at longer distances, and (2) self interaction needs to be avoided in the smallest simulation box size used in this study ($\sim$17 {\AA}). Within the cutoff, there could be some unphysical fluctuations in the force field (weights) because smoothness is not enforced in the ML method. To enhance the smoothness in resulting tabulated potentials, we apply either of the two simple post-processing modifications as described below. 

In the first option, the force field obtained from the ML model are fitted with specific function forms. Note that this is a simple curve fitting exercise and no MD simulations are needed for this step. We test three commonly used forms for interactions in the short range ($r<r_{cut}^{SR}$), i.e., Buckingham,\cite{buckinghamClassicalEquationState1938} Lennard-Jones (LJ), and Morse potentials with explicit Coulomb interactions,

\begin{equation}
    F^{Buck-Coul}(r_{ab})=A_{ab} B_{ab}\exp(-B_{ab}r_{ab})-6\frac{C_{ab}}{r_{ab}^7} + \frac{q_a q_b}{r_{ab}^2}
\end{equation}

\begin{equation}
    F^{LJ-Coul}(r_{ab}) = 24 \varepsilon [2(\frac{\sigma}{r})^{13}-(\frac{\sigma}{r})^7] + \frac{q_a q_b}{r_{ab}^2}
\end{equation}

\begin{align}
    F^{Morse-Coul}(r_{ab}) = & 2 \alpha D_e e^{-\alpha (r - r_0)} \left(1 - e^{-\alpha (r - r_0)}\right) \notag \\ & + \frac{q_a q_b}{r_{ab}^2}
\end{align}
% and find that the Coulomb–Buckingham function might be the most suitable for the short-range interactions ($r<r_{cut}^{SR}$),
where $a$ and $b$ denote two atomic species, ($A$, $B$, $C$) in $F^{Buck-Coul}$, ($\sigma$, $\varepsilon$) in $F^{LJ-Coul}$, and ($D_e$, $\alpha$, $r_0$) in $F^{Morse-Coul}$ are all fitting parameters. The basic constants $1/4 \pi \epsilon_0$ in Coulomb forces are included but not written for simplicity. The charges $q_{a,b}$ (for each form separately) are determined by the fitting process as well under the constraint of $q_\mathrm{Si} = -2q_\mathrm{O}$. 
Because the force field has already been revealed from ML, fitting these parameters can be straightforward by optimizing a loss function. Considering the fact that the number of pairs increases with distances in the average order of $r^2$ and the errors (of the same magnitude) in the force fields are more detrimental if at large distances, we particularly emphasize the large distance errors in the fitting loss function, 

\begin{equation}\label{eq:fit}
    \mathcal{L}_\mathrm{fit} = \sum_{r_{ab,i}} r_{ab,i}^2 \cdot |f^\mathrm{fit}(r_{ab,i})-f^\mathrm{FM-DFT}(r_{ab,i})|^2
\end{equation}
where $r_{ab,i}$ ranges from the distance between the nearest neighbor for the specific pair ($a,b$) to the short range cutoff $r_{cut}^{SR}$. 
For the long-range part ($r_{cut}^{SR}<r<r_{cut}^{LR}$), we choose the Wolf truncation method for efficiency,\cite{wolfExactMethodSimulation1999}
\begin{equation}
    V^{\mathrm{Wolf}}(r_{ab})=q_a q_b \left( \frac{1}{r_{ab}}-\frac{1}{r_{cut}^{LR}}+\frac{r_{ab}-r_{cut}^{LR}}{(r_{cut}^{LR})^2} \right)
\end{equation}
as followed by previous silica potential developments.\cite{,carreAmorphousSilicaModeled2007,sundararamanNewOptimizationScheme2018} For the boundary between SR and LR interaction, we multiply the above potentials with a commonly-used window function,\cite{carreDevelopingEmpiricalPotentials2016}
\begin{equation}
    G(r) = \exp \left( -\frac{\gamma^2}{(r-r_{cut}^{SR})^2} \right)
\end{equation}
where $\gamma$=0.2 {\AA} controls the width of the smoothing function.

In the second option for enhancing smoothness, we smooth the force fields in the short range by applying the Savitzky-Golay filter with a polyorder of 5 and window length of 2 {\AA}. Here we choose $r_{cut}^{SR}$=7.98 {\AA} to make the smooth range as wide as possible, and apply the Wolf truncation for the LR (only 0.02 {\AA} in this case) to ensure forces decay to zero at $r_{cut}^{LR}$. In addition, we can also enforce smoothness by adding additional regularization terms into the loss function based on roughness of the force field. However, the solution needs to be searched by gradient descent instead of simple matrix multiplication using Ridge.

Nevertheless, the method proposed here is not limited to any choice of function forms we listed in this subsection. One can apply any mathematical forms as needed or simply use the smoothed tabular force fields. 

\subsection{Learning data preparation}

As detailed in the earlier section, a Ridge regression ML model is utilized to obtain a force field that is the closest approximation of the quantum mechanical force field within the pair approximation. The descriptors $X_{aI}[b,i]=\sum_{bJ_i} \hat{u}(\mathbf{R}_{aI,bJ_i})\cdot \hat{x}$ 
% ($a$, $b$ denote two species, and $I$, $J$ are atomic indices belonging to the corresponding species) 
can be calculated easily from a given local configuration of atoms and are related to both the number of neighboring atoms for atom $I$ and the relative atomic positions in the shell. The interval between adjacent shells is set 0.02 {\AA}, resulting in a total dimension of 1200 for all the input feartures. The output $F_{I,x}^\mathrm{QM}$ are directly collected from ab initio calculations. For developing force fields involving multiple species, we can obtain $f_{ab}(r)$ by either learning forces on $a$ atoms or on $b$ atoms, the results of which may have slight differences for numerical reasons. A trick to prevent this is to train the models on different species altogether with all pairs of $X$. For instance, the final input for silica contains all Si-Si, Si-O, and O-O for each atom in one data structure. For Si atoms, $X_\mathrm{OO}$ are set to zeros and vice versa for O atoms. By using the entire data structure in training, a single force field involving different species, i.e., $f_{ab}(r)=f_{ba}(r)$, can be obtained. 

In this study, amorphous configurations for training this ML model are collected from ab initio molecular dynamics (AIMD) simulations based on DFT, implemented in Vienna Ab initio Simulation Package (VASP).\cite{kresseEfficiencyAbinitioTotal1996} The generalized gradient approximation of Perdew-Burke-Ernzerhof is used for the exchange-correlation functional.\cite{perdewGeneralizedGradientApproximation1996} Corrections for the van der Waals interactions are made using the DFT-D3 method with Becke-Johnson damping.\cite{grimmeEffectDampingFunction2011} In all the calculations, the electronic wave
function is expanded using a plane wave basis up to an energy
cutoff of 500 eV. The convergence criteria for the energy during self-consistent field calculations is set to 10\textsuperscript{-5} eV. 

A cubic simulation box containing 450 atoms, i.e., 150 SiO\textsubscript{2} units, is utilized in this study. Different densities ranging from 1.5 to 2.8 g/cm\textsuperscript{3} are covered for better mapping the energy landscape of interest, resulting in box lengths varying from 17.48 - 21.53 {\AA}. Due to the relatively large simulation system, only the gamma point in the reciprocal space is considered. The timestep is set to 1 fs. The temperature varies from 3000 - 4000 K, allowing for adequate sampling of different disordered structures within a relatively short trajectory.

Since the ML model is focused on forces on individual atoms and their local environments, each atom in a snapshot of the AIMD trajectory is a data instance. As a result, more than 400,000 data instances can be efficiently generated from a 300 fs trajectory, which can be completed in about one day of DFT calculation using 20 CPUs.

\section{Results}
\subsection{Predicting DFT forces by machine learning}

\begin{figure}[h]
  \centering
  \includegraphics[width=0.9\linewidth]{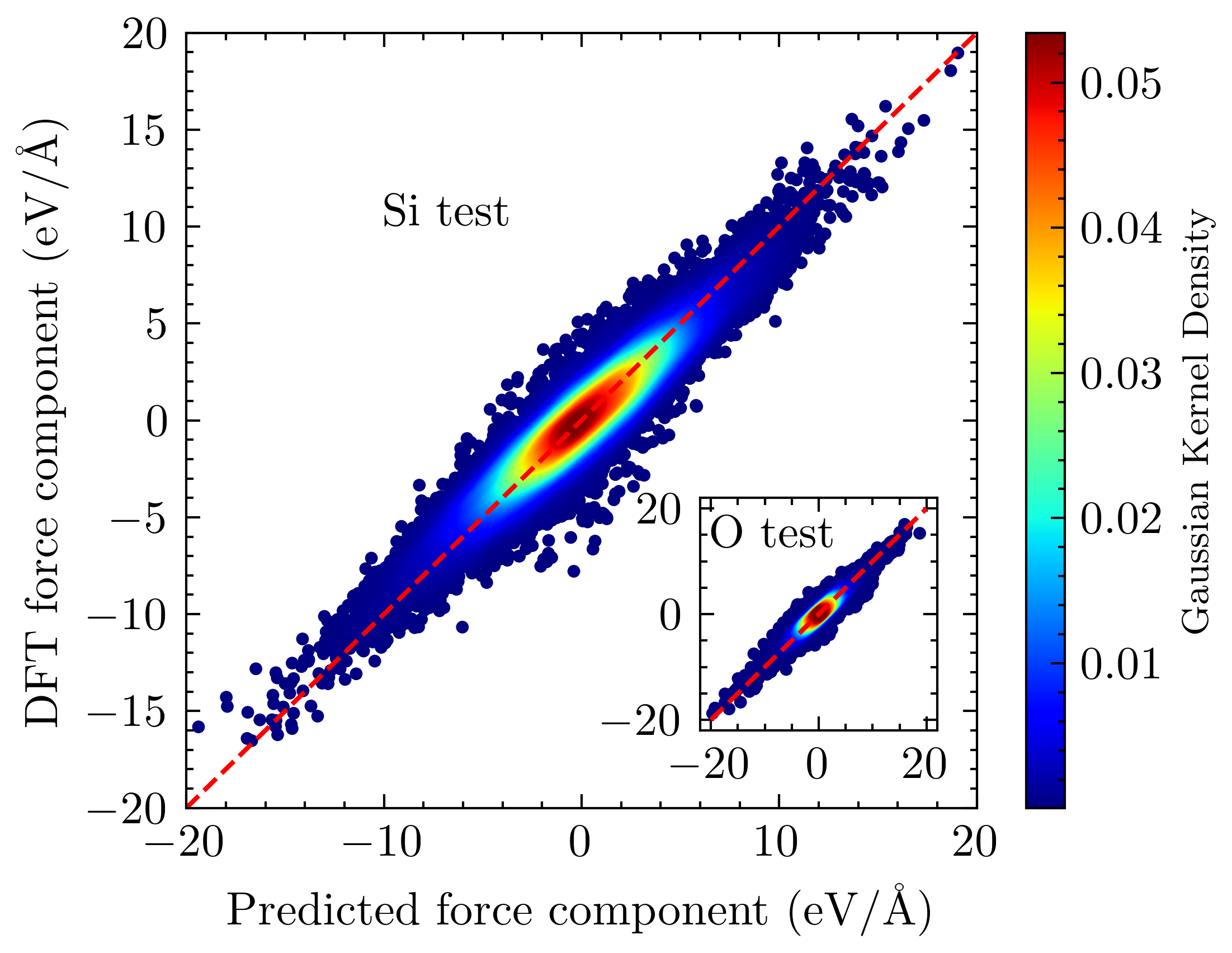}
  \caption{Comparison of the force components (forces projected onto $x$, $y$, and $z$ axes) predicted by the ML model with those calculated by DFT on the testing Si and O (inset) data. The dashed line representing the perfect match is a guide to the eye. The brighter colors representing higher data density lie closely to the dashed line, suggesting that the majority of forces are well predicted by the ML model.} %(b) R2 score for test data in systems with different sizes. Note that the ML model was trained in the system containing 450 atoms, i.e., `N450'.
  \label{fig:performance}
\end{figure}

Figure \ref{fig:performance} presents the performance of the force matching model for silica, validated against test data. The force components on individual atoms predicted by the model align well with those calculated by DFT, observed for both Si and O atoms. Notably, small forces stemming from ordered local structures exhibit larger deviations compared to large forces arising from distorted local structures. The RMSE values for Si and O test data are 1.14 and 0.72 eV/{\AA} respectively, resulting in an overall RMSE of 0.90 eV/{\AA}. Efforts to refine the learning outcomes via the expansion of the training data pool did not yield better results, as shown in the Supplementary Material. Further improvement would necessitate the incorporation of additional degrees of freedom, such as the inclusion of 3/4-body interactions or the deployment of more complex ML architectures.

\begin{figure}[h]
  \centering
  \includegraphics[width=1\linewidth]{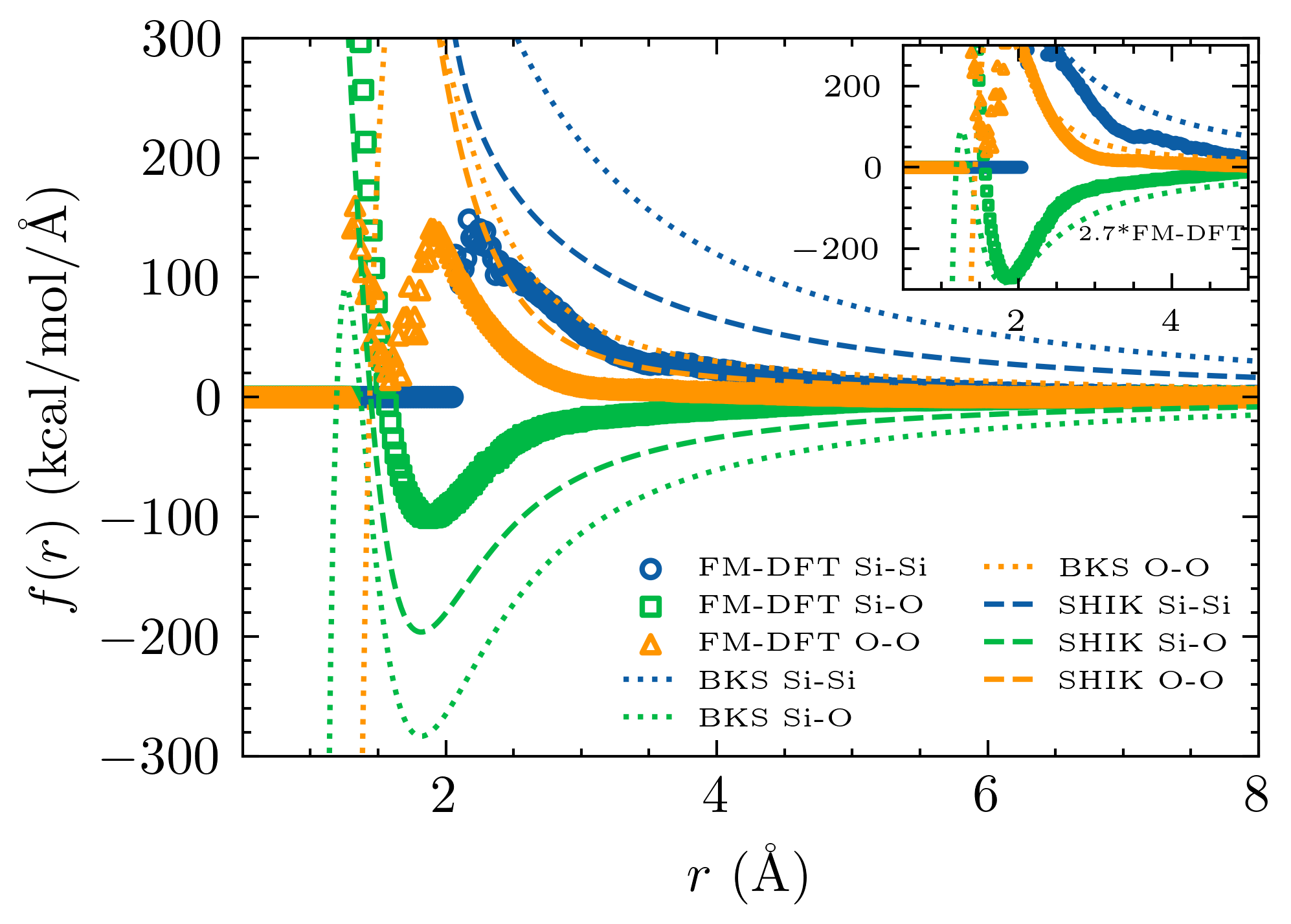}
  \caption{Pair force fields obtained from the ML model (denoted as `FM-DFT') as functions of pair separation, compared to the BKS and SHIK potential. The inset shows that the FM-DFT force fields have different shapes compared to the BKS potential apart from the scale (`FM-DFT' is rescaled by 2.7 times in the inset).}
  \label{fig:force}
\end{figure}

As detailed in Sec. \ref{sec:method_ML}, the weights of the linear regression model based on Equation 6 directly represent a force field as a function of distance. These force fields, henceforth referred to as `force-match-DFT' (FM-DFT), display notable differences from SHIK and BKS, as illustrated in Fig. \ref{fig:force}. Specifically, FM-DFT forces are about 2-3 times smaller for Si-O around 2 {\AA}. Furthermore, the force profiles vary significantly, although the attraction minima of Si-O forces for all three potentials are similarly positioned. Notably, FM-DFT forces exhibit a more rapid decline within 2-3 {\AA} compared to both BKS and SHIK, pointing to a softer medium-range order (MRO) interaction. Minor differences in short-range order (SRO) are also evident, with the zero point of Si-O FM-DFT forces slightly offset.

We also notice some unphysical forces at very short ranges (far shorter than the nearest neighbor distance), probably resulting from sparse data in these areas. Specifically, in situations where no atomic pair exist at a particular separation, the force field default to zero due to regularization. When only a small number of atomic pairs are available, large variances exhibit in the generated force field, such as those observed around 1.5-2 {\AA} for O-O. These deviations at very short ranges, however, do not impact simulations as long as pressure or temperature are not extremely high (e.g., $>$6000 K).\cite{vollmayrCoolingrateEffectsAmorphous1996} Therefore, we simply apply a linear force fit (harmonic approximation) in the affected range to prevent unphysical collision, with cutoffs detailed in Table S1 in the Supplementary Material.

Within the effective range of potentials ($r_{cut}^{linear}<r<r_{cut}^{LR}$), small force variations might also occur due to the finite resolution of discrete distances. As elaborated in Sec. \ref{sec:potentialsetting}, we explored two options to enhance potential smoothness, 1) function-form fitting, or 2) direct smoothing. 

\begin{figure}[h]
  \centering
  \includegraphics[width=0.9\linewidth]{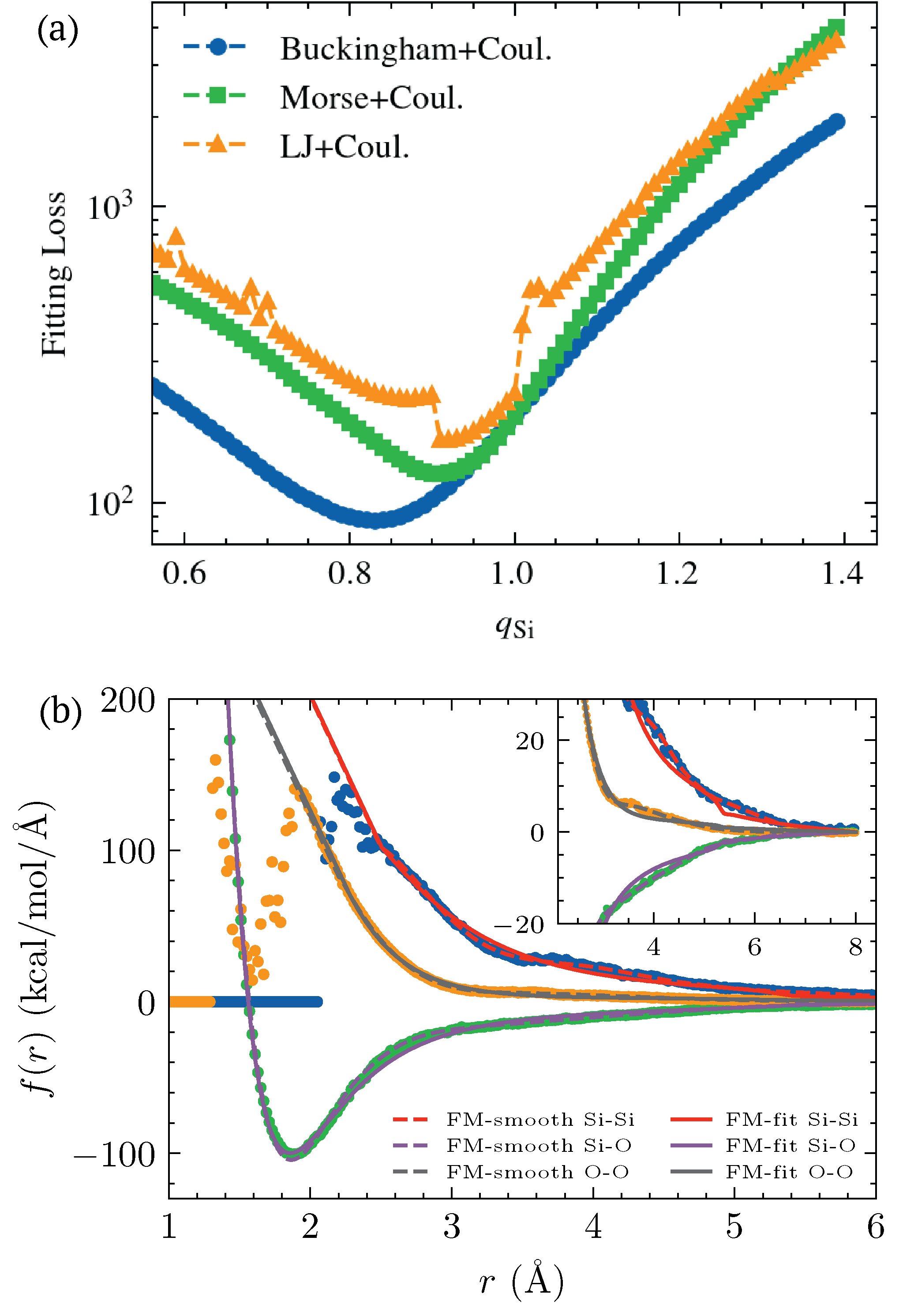}
  \caption{(a) Fitting loss defined in Equation \ref{eq:fit} as a function of the charge of Si for the three fitting function forms as detailed in Sec. \ref{sec:potentialsetting}. The Buckingham-Coulomb form with $q_\mathrm{Si}=0.82$ has the lowest fitting loss from `FM-DFT', and therefore, is selected for further validation, which will be denoted as `FM-fit' hereinafter.  
  (b) Comparison between the force fields generated by fitting and smoothing after obtaining the raw force field from the ML model `FM-DFT', as detailed in Sec. \ref{sec:potentialsetting}. The inset shows the subtle changes at the long range.}
  \label{fig:ff_compare}
\end{figure}

Figure \ref{fig:ff_compare}a summaries the fitting loss defined in Equation 10 against the charge of Si for three commonly used function forms: Buckingham, LJ, and Morse, all combined with Coulomb interaction. The Buckingham-Coulomb form with $q_\mathrm{Si}=0.82$ exhibits the lowest fitting loss. This combination, henceforth referred to as `FM-fit', aligns remarkably with `FM-DFT', as shown in Fig. \ref{fig:ff_compare}b. In contrast, both the LJ-Coulomb and Morse-Coulomb models falter in accurately capturing short-range interactions, as shown in Supplementary Material. This observation accentuates the Buckingham-Coulomb interaction's capability to best represent the effective pair interaction in silica, confirming previous preferences in classical force fields for silica. However, the effective charge fitted from the `FM-DFT' $q_\mathrm{Si}=0.82$ is smaller than previously used values (2.40 and 1.74 in BKS and SHIK, respectively), suggesting a stronger Coulomb screening effect than expected. The parameters of `FM-fit' are summarized in Table S1 in the Supplementary Material.

The force field directly directly smoothed from `FM-DFT', denoted as `FM-smooth', reveals very subtle differences from `FM-fit', as shown in Fig. \ref{fig:ff_compare}. To determine their actual effectiveness in silica simulations, it's essential to conduct tests in various MD situations. 

\subsection{Performance of the generated force field}
\subsubsection{Melt-quenching behaviors}\label{sec:mq}

\begin{figure}[h]
  \centering
  \includegraphics[width=0.9\linewidth]{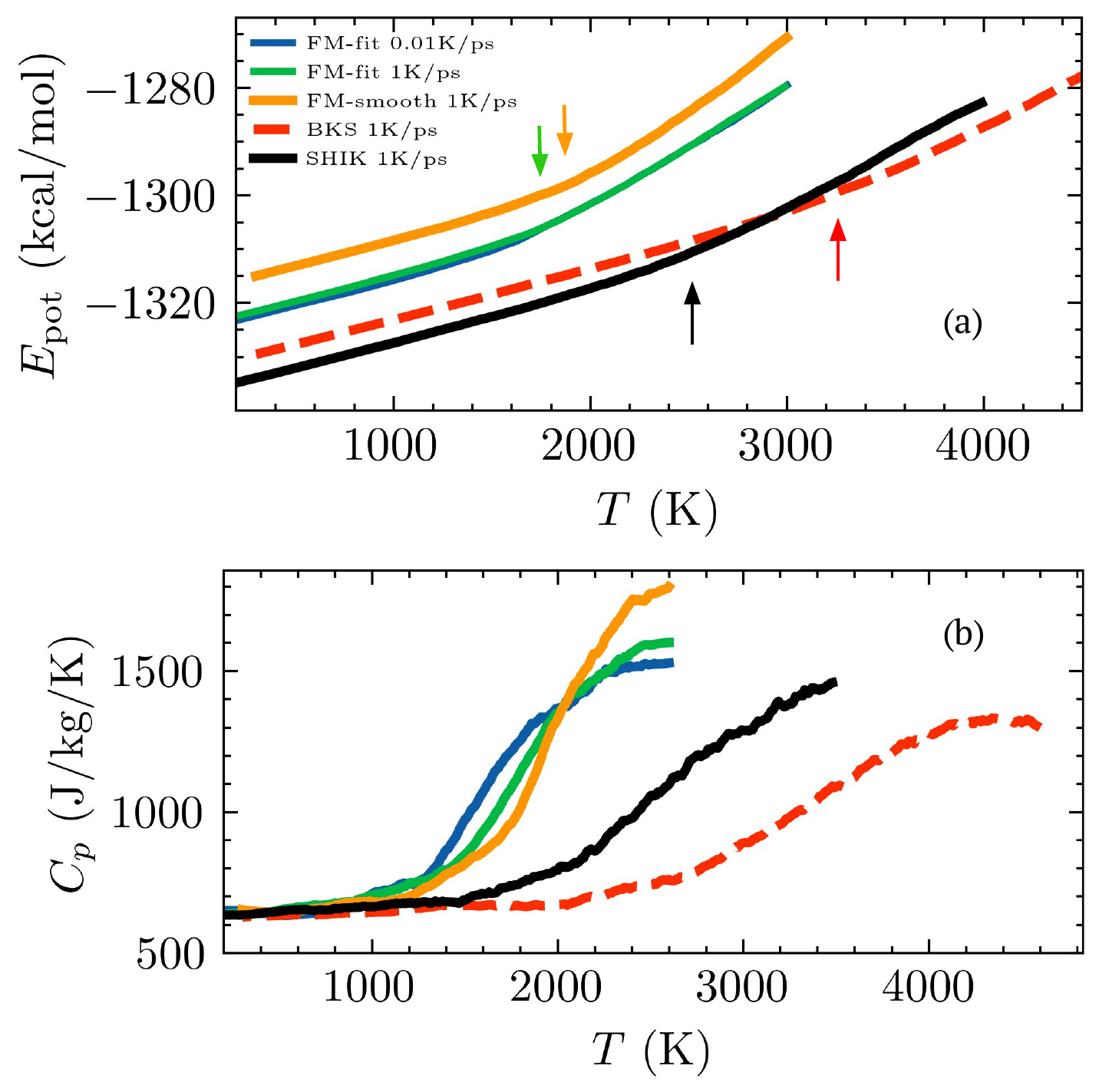}
  \caption{(a) Potential energy and (b) heat capacity after removing the thermal fluctuation as a function of temperature for silica simulated by different classical potentials. Energies in (a) have been shifted vertically for easier comparison. The glass transition temperature obtained by the FM potentials, as marked by arrows in (a), are significantly lower than those by the SHIK and BKS potentials.}
  \label{fig:e_cooling}
\end{figure}

The obtained potentials in the tabulated forms are then tested in MD with the Large-scale Atomic/Molecular Massively Parallel Simulator (LAMMPS).\cite{LAMMPS} Figure \ref{fig:e_cooling} illustrates the potential energy (a) and heat capacity (b) as a function of temperature for silica simulated by the FM potentials and commonly used empirical potentials (BKS and SHIK) during the melt-quenching simulations in the isothermal-isobaric (NPT) ensemble. Simulation settings with different potentials are all the same. All the results of the melt-quenching simulations hereinafter are average of at least five independent runs using the same setting but different initial atomic velocities. Although the glass transitions are all observed and the glass heat capacity are almost the same for these potentials, $T_g$ of the FM-DFT potentials are around 1800 K, as listed in Table \ref{tab:table2}, significantly lower than the SHIK (by $~$700 K) and BKS potentials (by $~$1400 K), but  still higher than the experimental $T_g$ due to the large cooling rate. This suggests dynamics of supercooled liquids simulated by the FM-DFT potentials is faster by orders of magnitude, which is probably due to the softening of potentials as shown in Fig. \ref{fig:force}. Note that the melting temperature of silica in experiments is around 2000 K,\cite{haynesCRCHandbookChemistry2014} indicating that $T_g$ given by the FM potentials are more reasonable. The potential energy of silica glass and the $T_g$ are lower if cooled with a smaller cooling rate, consistent with the glass transition expectation. In addition, the silica liquids simulated by the FM potentials have slightly higher specific heat capacity than that of the BKS and SHIK potentials. %, which are closer to the experimental value.\cite{}

\begin{table}[h]%The best place to locate the table environment is directly after its first reference in text
\caption{\label{tab:table2}%
Properties of silica glasses simulated by different potentials, including $T_g$, heat capacity and density at 300 K, and Young's modulus at 0 K. A constant cooling rate of 1K/ps is used in the melt-quenching simulations. The BKS potential used here is applied with a SR cutoff of 6 {\AA}.}
\begin{ruledtabular}
\begin{tabular}{lcccc}
& \shortstack{$T_g$ \\ (K)} & \shortstack{$C_p$ \\ (J/Kg/K)} & \shortstack{Density \\ (g/cm\textsuperscript{3})} & \shortstack{ E \\ (GPa)} \\
\colrule
FM-fit & 1760 & 641  & 2.27 & 60.53 \\
FM-smooth & 1832 & 653  & 2.29 & 52.55\\
BKS & 3232 & 628  & 2.31 & 88.30\\
SHIK & 2533 & 635 & 2.13 & 57.23\\
Exp.\cite{PropertiesFusedSilica,wiki:Fused_quartz} & 1475-1480\cite{richetGlassTransitionsThermodynamic1984} & 680-730 & 2.20 & 70-73\\
\end{tabular}
\end{ruledtabular}
\end{table}

\begin{figure}[h]
  \centering
  \includegraphics[width=0.9\linewidth]{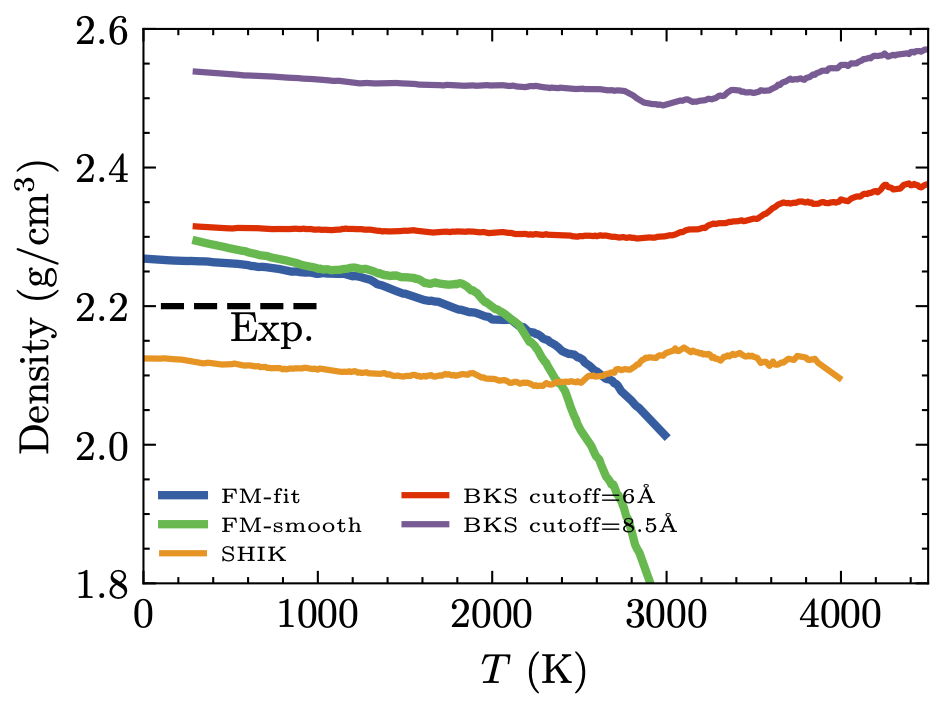}
  \caption{Density as a function of temperature during melt-quenching simulations of silica simulated by different potentials.}
  \label{fig:density_cooling}
\end{figure}

Density changes during the melt-quenching simulations by different potentials are shown in Fig. \ref{fig:density_cooling}. In general, the densities of glasses obtained by these potentials are not far from the experimental value except the BKS potential with a relatively large cutoff like 8.5 {\AA}. However, large differences arise in the liquid region. The supercooled liquids simulated by the FM potentials have large positive thermal expansion coefficients, whereas those simulated by the BKS and SHIK potentials have small negative thermal expansion coefficients at least within $T_g<T<T_g$+1000 K. This leads to decreasing density of supercooled liquids for the FM potentials at higher temperatures.
%, which qualitatively agrees with the behaviors simulated by the SHIK potential in a previous study.\cite{sundararamanNewOptimizationScheme2018} 
Pushing the temperature beyond 3000 K for simulations employing the FM potentials leads to system instability (explosion), resonating with silica's experimental boiling point of approximately 2500 K.\cite{haynesCRCHandbookChemistry2014} These findings underscore considerable issues with earlier potentials, e.g., BKS and SHIK, when modeling silica liquids, and suggest that the previously reported abnormal relationship between density and thermodynamic stability could be exaggerated as consequences of flawed potential parameters.\cite{yuStructuralSignaturesThermodynamic2021} The FM potentials appear poised to rectify this issue in liquid simulations. 

\subsubsection{Liquid structures at high temperatures}

\begin{figure}[h]
  \centering
  \includegraphics[width=0.9\linewidth]{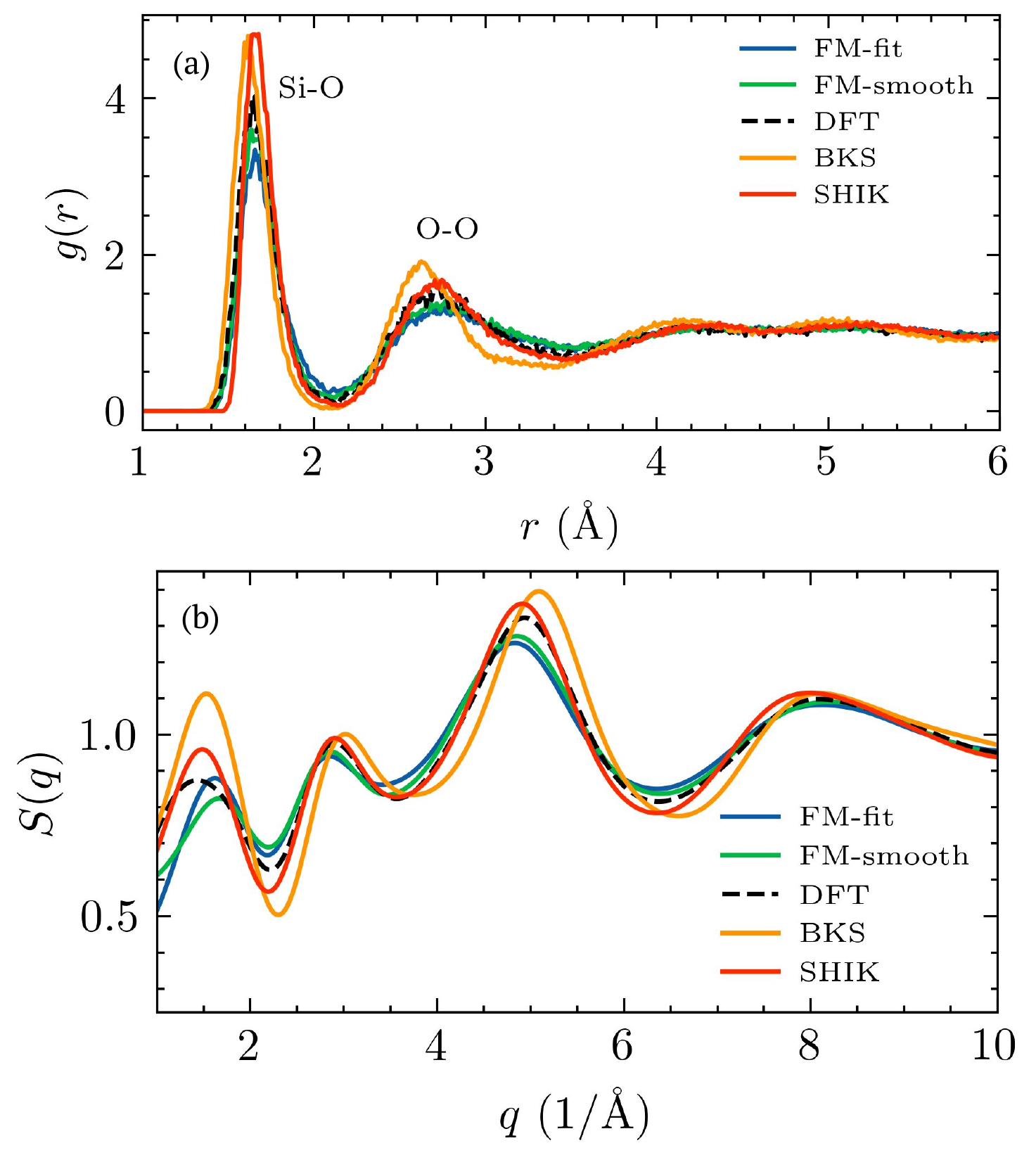}
  \caption{(a) Pair distribution function and (b) static structure factor of silica liquids at 3000 K simulated by different methods.}
  \label{fig:s1_hT}
\end{figure}

Next, we analyze the atomic structures of silica liquids simulated by the FM potentials. Figure \ref{fig:s1_hT} shows the pair distribution function $g(r)$ and static structure factor $S(q)$ of liquid structures in equilibrium at 3000 K simulated by different methods. The results of classical MD are averaged over 100 structures collected every 1 ps, and the DFT results are averaged over 40 structures collected every 5 fs from the production runs. In general,  the FM potentials are better than the BKS potential to reproduce the 2-point density correlation in AIMD (DFT) simulations of silica liquids, but does not show evident advantages over the SHIK potential.  In the short range, i.e., 1.6-2.2 {\AA} in $g(r)$, the peak locations of the FM potentials are in better agreement with the DFT results than SHIK, but the peak heights are lower due to softer interactions. In the medium-range regime, e.g., above 2.2 {\AA} in $g(r)$ or below 4 {\AA}$^{-1}$ in $S(q)$, the FM potentials can generally reproduce the structural characteristics but the simulated structures seem more amorphous than other methods. 

\begin{figure}[h]
  \centering
  \includegraphics[width=0.9\linewidth]{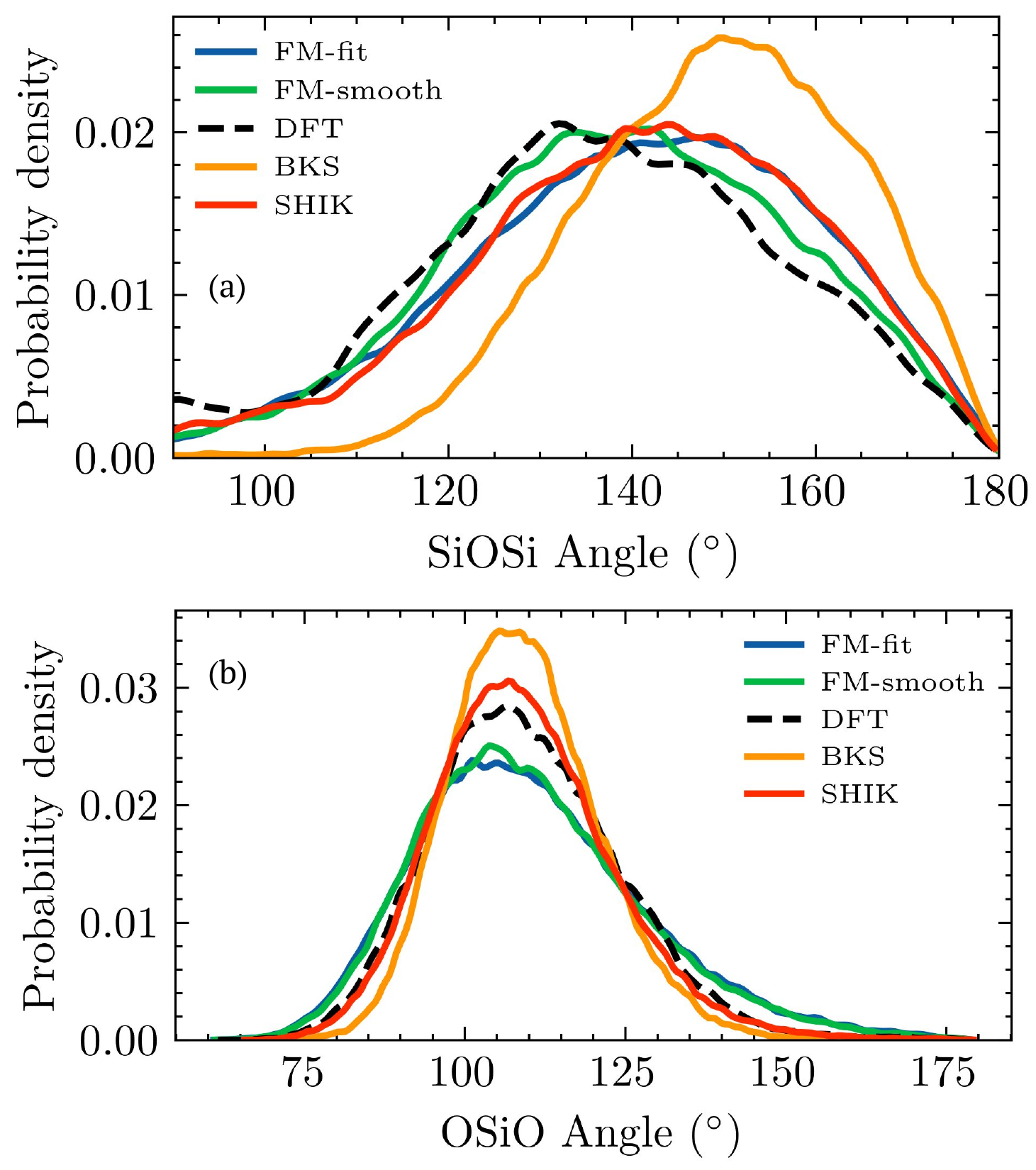}
  \caption{(a) Si-O-Si angle distribution and (b) O-Si-O angle distribution of silica liquids at 3000 K simulated by different methods. The structures are the same as those in Fig. \ref{fig:s1_hT}.}
  \label{fig:s2_hT}
\end{figure}

Figure \ref{fig:s2_hT} shows the Si-O-Si and the O-Si-O angle distributions of silica liquid structures at 3000 K simulated by different methods. Again, the BKS structures deviate the most from the DFT structures, especially in the Si-O-Si angle. The FM potentials perform similarly to the SHIK potential in the Si-O-Si angles but show wider distributions along with lower peak heights in the O-Si-O distribution, suggesting the FM potentials allow slightly larger variations of SiO\textsubscript{4} tetrahedra. This is again a result of the softer interactions in the short range. Nevertheless, we suppose that the FM potentials are able to reproduce reasonable liquid structures of silica. Comparison of glass structures simulated by the FM potentials and other methods are included in SI. Note that the simulated glass structures are not supposed to match perfectly with the experimental structure since the cooling rate difference is around 10 orders. 

\subsubsection{Cooling rate effects of mechanical properties}

\begin{figure}[h]
  \centering
  \includegraphics[width=0.9\linewidth]{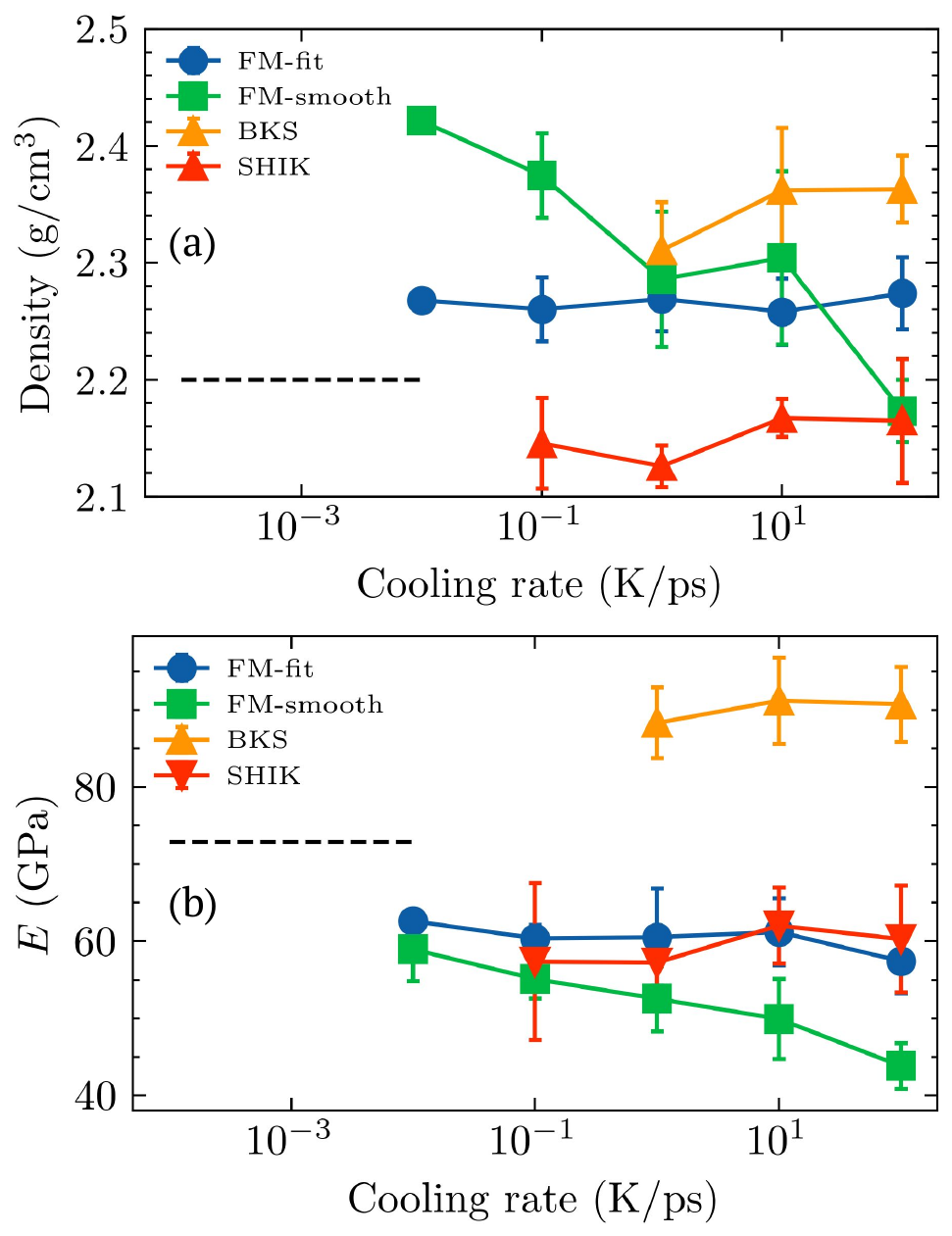}
  \caption{Cooling rate effects on (a) density and (b) Young's modulus of silica glasses simulated by different classical potentials. The dashed lines denote the experimental values with a much smaller cooling rate $\sim$1 K/s.}
  \label{fig:cooling_rate}
\end{figure}

A key aspect of glassy materials is that the materials properties closely depend on the thermal history, such as, the cooling rate during the melt-quenching process. Therefore, we also evaluate the cooling rate effects of density and Young's modulus for the FM potentials, as shown in Fig. \ref{fig:cooling_rate}. The cooling rates vary from 100 to 0.01 K/ps in this study, and the results are average of at least five independent melt-quenching runs. Although the two FM potentials are similar (Fig. \ref{fig:ff_compare}), they behave differently in the cooling rates effects. FM-smooth shows a strong cooling rate dependence, i.e., the glass becomes denser and stronger more quickly when lowering the cooling rate, which however leads to an overshoot in density compared to experiments. FM-fit however shows a mild cooling rate dependence, from which the density is almost constant (2.27 g/cm$^3$) and the Young's modulus increases slowly when cooling slowly, the extrapolation of which might be very close to the experimental value.

For the traditional potentials BKS and SHIK, they both show negative cooling rate effects, i.e., the glass becomes less dense and mechanically weaker when lowering the cooling rate, which contradicts the general trend for almost all other glasses. This abnormal cooling rate effect is likely a consequence of the problem of liquid simulations for the two potentials, as earlier shown in Fig. \ref{fig:density_cooling}. The previous studies based on these potentials might not reveal real silica's behaviors.\cite{vollmayrCoolingrateEffectsAmorphous1996,horbachStaticDynamicProperties1999} In addition, this raises further question on the dynamics of liquid silica such that whether the well studied fragile-to-strong transition (reported by the BKS simulations) occurs at a lower temperature or does not exist at all in the real material.\cite{saksaengwijitOriginFragiletoStrongCrossover2004,yuUnderstandingFragiletoStrongTransition2022,horbachStaticDynamicProperties1999} 

\subsection{Physical insights from applications in silicates}

The force-match methodology introduced in this study offers more than just a tool for crafting classical potentials. It furnishes a direct, user-friendly avenue for probing variations in atomic interactions under diverse conditions. One pertinent observation stems from analyzing silica force fields derived from configurations spanning densities from 1.5 to 2.8 g/cm$^3$. The distinctions among these force fields are subtle, staying within 1 eV/{\AA} across all distances, as shown in Fig. S3 in the Supplementary Material. Such minimal discrepancies suggest that silica's atomic interactions are only marginally influenced by density. Thus, a singular force field might suffice for simulating silica across various pressures while maintaining satisfactory precision.

\begin{figure}[h]
  \centering
  \includegraphics[width=0.9\linewidth]{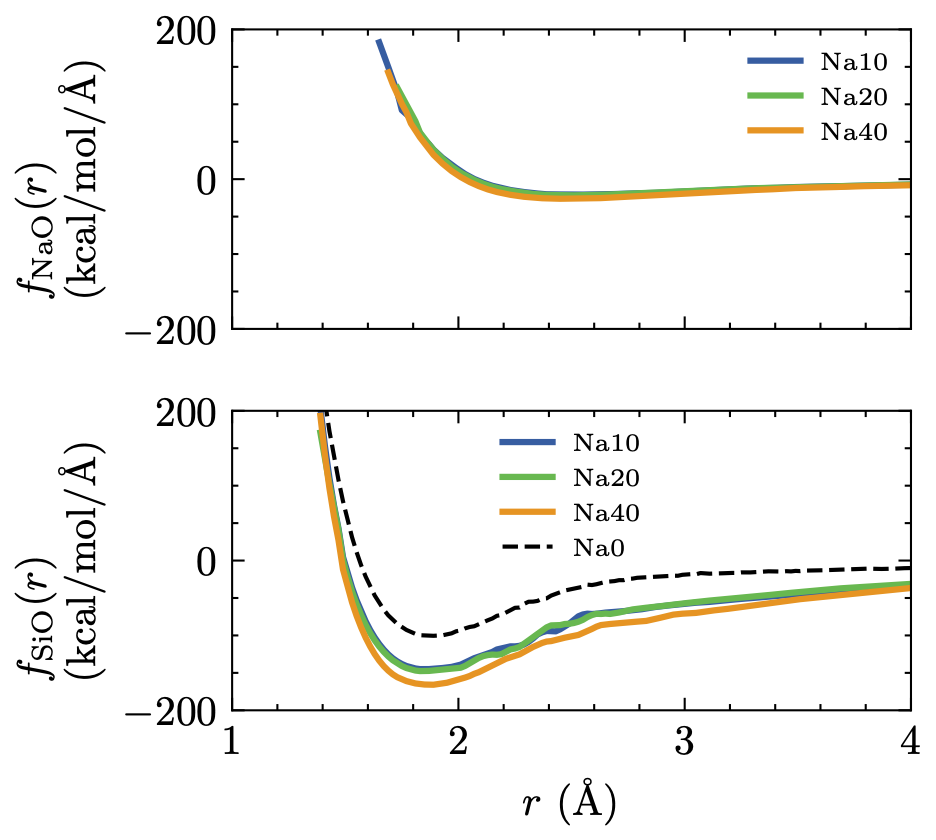}
  % \caption{ Force fields of (a) Na-O and (b) Si-O as functions of distances at different Na concentrations in sodium silicate (NaO)$_x \cdot$(SiO$_2$)$_{1-x}$, where $x$ varies from 0 to 40.  The Si-O forces in sodium silicate change significantly with the sodium concentration, suggesting that a composition-dependent force field is needed to accurately simulate sodium silicate.}
  \caption{Force fields of (a) Na-O and (b) Si-O mapped as functions of distances at various sodium concentrations in sodium silicate, represented as (NaO)$_x \cdot$(SiO$_2$)$_{1-x}$ where $x$ ranges from 0 to 40. The notable shift in Si-O forces in sodium silicate with sodium concentration underscores the need for a composition-dependent force field for accurate simulations.}
  \label{fig:potential_silicate}
\end{figure}

We also extend the approach to evaluate force fields in sodium silicate and borosilicate with different compositions. The similar AIMD simulation settings used for silica were employed to gather training data for the linear ML models. As detailed in Supplementary Material, predicted forces in sodium silicate and borosilicate align well with their DFT counterparts. The forces on sodium (Na) atoms exhibit even higher accuracy in predictions compared to those on silicon (Si) and oxygen (O) due to simpler coordination environment. 
Our approach allows for an uncomplicated assessment of pair interactions' dependency on chemical environments, notably the concentration variations of specific species in multi-component systems. For instance, in sodium silicate, while Na-O forces remain consistent, the Si-O attractions strengthen as sodium concentration rises, as illustrated in Fig. \ref{fig:potential_silicate}. Note that these alterations are more pronounced than those observed due to changes in silica density. This shift could potentially be attributed to alterations in the concentration of bridging and non-bridging oxygen. Therefore, those looking to simulate sodium silicate using classical pair potentials should consider adopting a composition-dependent parameter set to obtain high simulation accuracy. Conversely, borosilicate interactions exhibit only minor shifts with changing boron concentrations, hinting that a singular parameter set, akin to that for silica, may suffice.

\section{Discussion}

In summary, this work proposes a new force-matching approach that can proficiently extract effective pair atomic interactions from ab initio calculations, leveraging regularized linear regression models. This method introduces a simple yet effective representation of local atomic environments, which quantifies the `number' of atom pairs projected onto a single axis. As a direct upshot of this  descriptor set, the regression model's weights encapsulate the target force field as functions of pair distance, which are closest to the quantum mechanical many-body interactions within the confines of the pair approximation. 

This research provides two applications of this method. Firstly, it fosters the development of classical force fields for systems that can be succinctly described by the pair approximation, exemplified by silica and silicates. We reiterate that although rooted in ML techniques, our approach distinctly diverges from ML potentials as well as conventional or ML-based fitting of classical potential parameters. Our methodology derives tabulated potentials or parameter sets that are synonymous in terms of format with established classical potentials. There's no need for subsequent ML model computations or descriptor evaluations post-development. As such, this streamlined approach surpasses ML potentials in efficiency by at least an order of magnitude. Simultaneously, it fully utilizes the force information from ab initio simulations of complex material structures to achieve higher fidelity over classical pair potentials fitted from configurational energy or observables (i.e., derivatives of energy). Notably, our approach does not use pre-defined function forms and produce fundamental information about the atomic interactions in the system.

Another notable feature of our method is the expediency in developing classical potentials, which is evident in our trials where we generated a new classical potential by a single day's worth of computation on a 20-CPU computer. Although our force field may lack the generalizability of some ML potentials, its swift development and integration stand out. It also serves as an advantageous precursor for top-down classical potential development, such as aligning with empirical thermodynamic data, significantly curtailing time spent on high-dimensional parameter space optimization. Further, our methodology boasts an enhanced extrapolative capability over ML potentials. Illustratively, while our training for the silica model currently focuses on liquid structures, the resultant potentials competently model glass structures and their inherent properties, even at lower temperatures. The region for extrapolability here is physically intuitive, i.e., among systems and conditions where interactions at the atomic level are not expected to change significantly. The examination of the extrapolable region can be achieved within the method as well, which leads to its second application. 

The second application of our force-matching approach revolves around enabling rapid and precise assessments of thermodynamically averaged atomic interactions within complex atomic environments, grounded in quantum mechanical calculations. Historically, there has been a knowledge gap concerning the accuracy with which prior empirical potentials of diverse functional forms described atomic interactions. For instance, for decades, classical force fields for silica and silicate glasses have selected the potential function form and compositional dependent or independent parameters with little justification. Our method bridges this gap by offering insights derived directly from structures, a departure from evaluations based on isolated forces from individual particles--in this study, we provide direct evidence supporting the use of the Coulomb–Buckingham function form in amorphous silica structures. Furthermore, by focusing on learning within structures, we can discern variations in effective atomic interactions as the environments and conditions shift. Such knowledge underlines the transferability of classical force fields. For instance, in silicate systems, our method efficiently detects changes in atomic interactions due to alterations in system variables like density or modifier concentrations (e.g., in sodium silicates and borosilicates), thereby aiding in the fine-tuning of theoretical or semiempirical models.

The methodology we introduce may, on the surface, appear constrained by the validity of the pair approximation for the subject material. However, delving into its foundational philosophy reveals its latent capacity to encompass higher-order interactions, such as angular forces, in a manner analogous to pairs. Future investigations might also extend our approach to multi-component systems or probe its alignment with more advanced quantum mechanical simulations. We believe that this physics-informed machine learning technique furnishes an alternative avenue, distinct from existing machine learning potentials, propelling a deeper comprehension and simulation of amorphous materials at the atomic level.

\section{Data Availability}
The force fields produced in this study, along with the corresponding codes for data collection and model training, are accessible for download at https://github.com/zyumse/FMpotential.

\begin{acknowledgments}
This research was primarily supported by NSF through the University of Wisconsin Materials Research Science and Engineering Center (DMR-1720415).This work used the Extreme Science and Engineering Discovery Environment (XSEDE), which is supported by National Science Foundation grant number ACI-1548562. 
\end{acknowledgments}

\bibliography{potential}% Produces the bibliography via BibTeX.

\end{document}

% \preprint{APS/123-QED}

\title{Supplemental Material of ``How close are the classical two-body potentials to ab initio calculations? Insights from linear machine learning based force matching"}% Force line breaks with \\

\author{Zheng Yu}
\affiliation{Department of Chemistry, University of Illinois at Urbana-Champaign, Urbana, 61801, USA}
%  \altaffiliation[Also at ]{Physics Department, XYZ University.}%Lines break automatically or can be forced with \\
\author{Ajay Annamareddy}
\affiliation{Department of Materials Science and Engineering, University of Wisconsin-Madison, Madison, 53706, USA}
\author{Dane Morgan}
\affiliation{Department of Materials Science and Engineering, University of Wisconsin-Madison, Madison, 53706, USA}
\author{Bu Wang}%
 \email{bu.wang@wisc.edu}
 \affiliation{Department of Materials Science and Engineering, University of Wisconsin-Madison, Madison, 53706, USA}
 \affiliation{%
 Department of Civil and Environmental Engineering, University of Wisconsin-Madison, Madison, 53706, USA
}%

\maketitle

%\tableofcontents

\section*{Contents}
\begin{itemize}
    \item Machine learning model convergence and testing
    \item Glass structures generated by the FM-DFT potentials
    \item FM-DFT force fields of silicates 
    % \item Machine learning potential energy
\end{itemize}

\newpage
% \section{Machine learning model convergence and testing}

%Figure \ref{fig:error_size} shows the training and testing error as a function of training data size. It can be seen that the two kinds of errors are converged and almost equal after training with more than $10^5$ data. 

\begin{figure}
  \centering
  \begin{minipage}[b]{0.48\linewidth}
        \includegraphics[width=\linewidth]{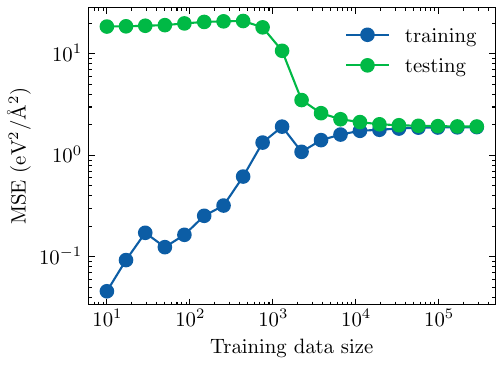}
    \end{minipage}
    \begin{minipage}[b]{0.48\linewidth}
        \includegraphics[width=\linewidth]{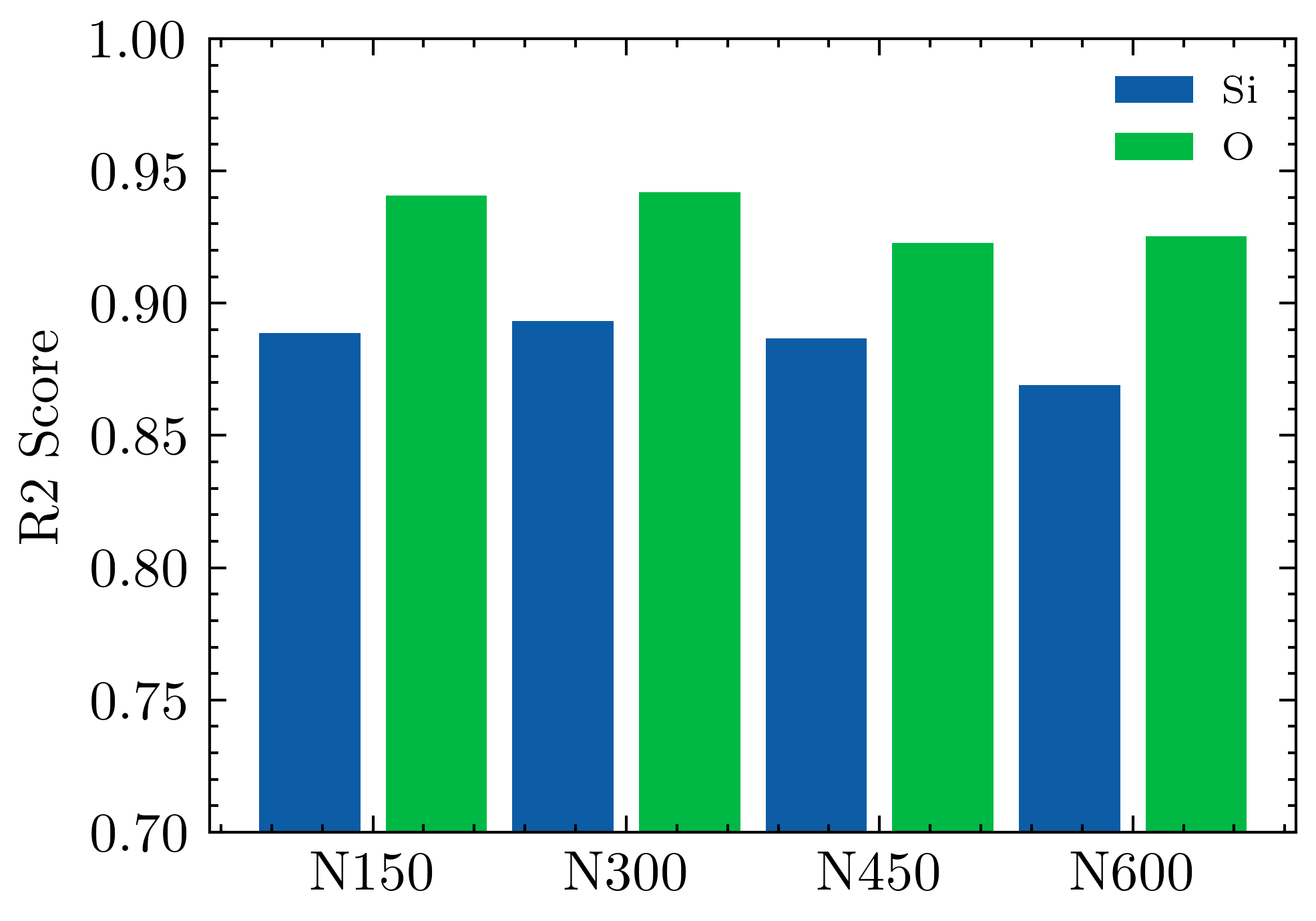}
    \end{minipage}
  \caption{(left) Training and testing error as a function of training data size averaged by five random selection from the entire training dataset. The bias$>$variance after 10$^4$ indicates the current dataset is large enough for training the model. Further improving the learning outcome requires introducing more complicated algorithms or more degrees of freedoms, e.g., adding multi-body interaction. (right) Consistent R2 score for test data in systems with different sizes. Note that the ML model was trained in the system containing 450 atoms, i.e., `N450'. }
  \label{fig:error_size}
\end{figure}

\begin{table}[h]%The best place to locate the table environment is directly after its first reference in text
\caption{\label{tab:table1}%
Parameters of the FM-fit potential in the form of Buckingham-Coulomb. $q_\mathrm{Si}$=0.82 and $q_\mathrm{O}$=$-q_\mathrm{Si}/2$. The SR cutoff $r_{cut}^{SR}$ =5.5 {\AA}, and the LR cutoff $r_{cut}^{LR}$ =8 {\AA}. }
\begin{ruledtabular}
\begin{tabular}{lccccc}
& $A$ (eV) & $B$ ({\AA}\textsuperscript{-1}) & $C$ (eV$\cdot${\AA}\textsuperscript{6}) & $r_{cut}^{linear}$ ({\AA})  \\ %& $k^{Near}$
\colrule
Si-Si & 302.498 & 1.883 & 226.827 & 2.4\\
Si-O & 30002.376 & 5.321 & 141.004 & 1.3\\
O-O & 1643.223 & 2.960 & 165.554 & 2.2\\
\end{tabular}
\end{ruledtabular}
\end{table}

\begin{figure}
  \centering
  \includegraphics[width=0.6\linewidth]{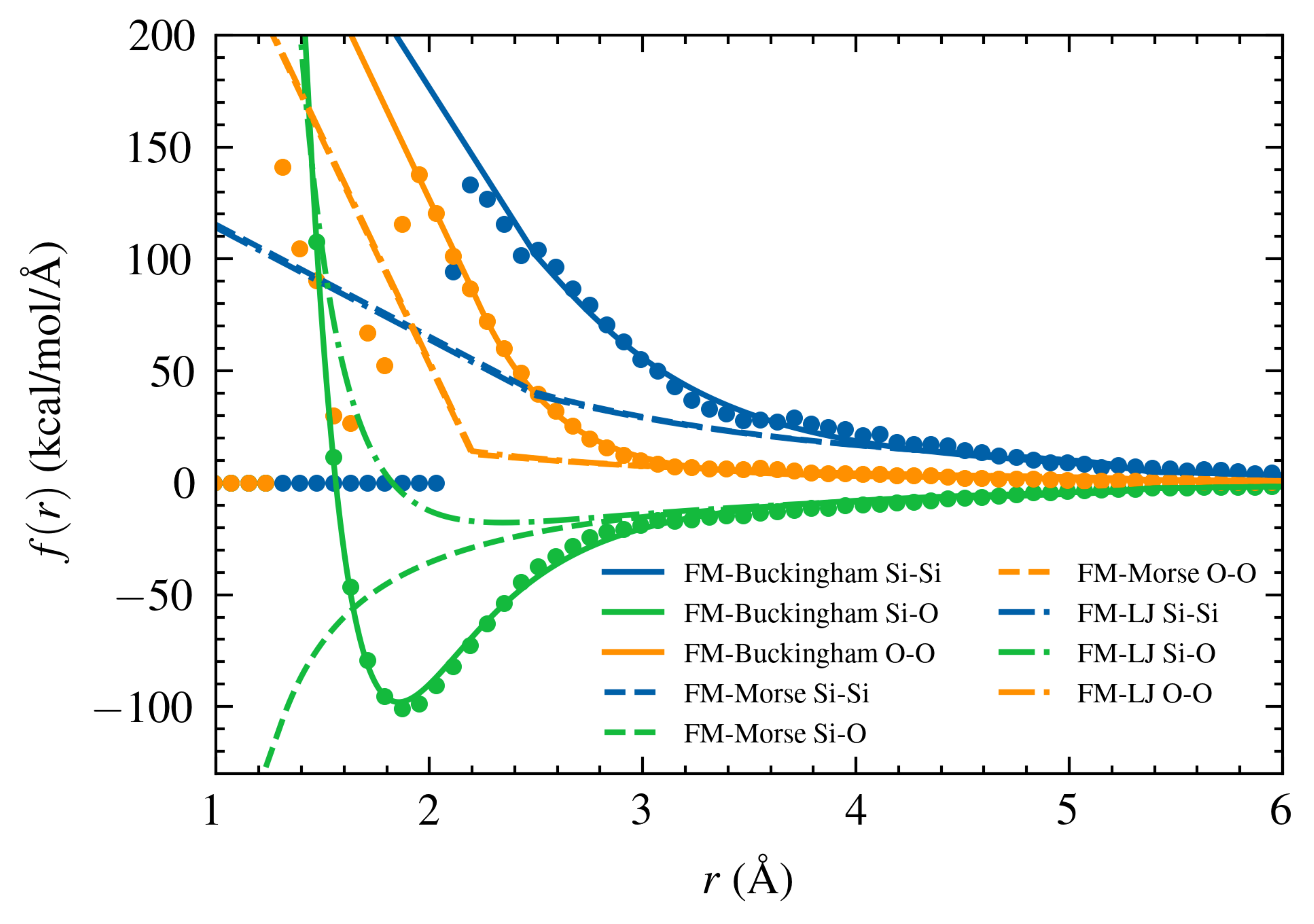}
  \caption{Force fields in the fitting forms of LJ-Coulomb and Morse-Coulomb cannot match the short-range interactions in `FM-DFT' as well as Buckingham-Coulomb, suggesting that Buckingham-Coulomb is the best empirical form to describe effective pair interactions in amorphous silica. }
  \label{fig:s1_lT}
\end{figure}

\begin{figure}
  \centering
  \includegraphics[width=0.6\linewidth]{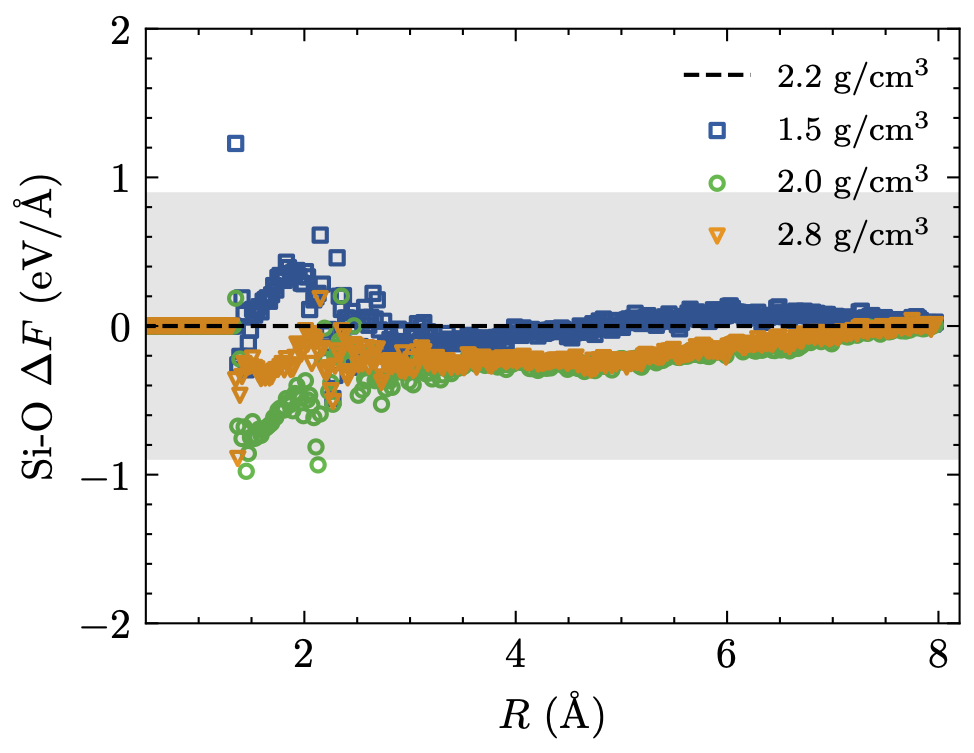}
  \caption{Changes of the Si-O force fields as functions of distance in silica with different densities. The reference is chosen as that with a density of 2.2 g/cm\textsuperscript{3}. The shadow regions shows the prediction error of the ML model.}
  \label{fig:s1_lT}
\end{figure}

% stability of weights (does not depend on density and force manipulation)

% \section{Glass structures generated by the FM-DFT potentials}
\begin{figure}
  \centering
  \includegraphics[width=\linewidth]{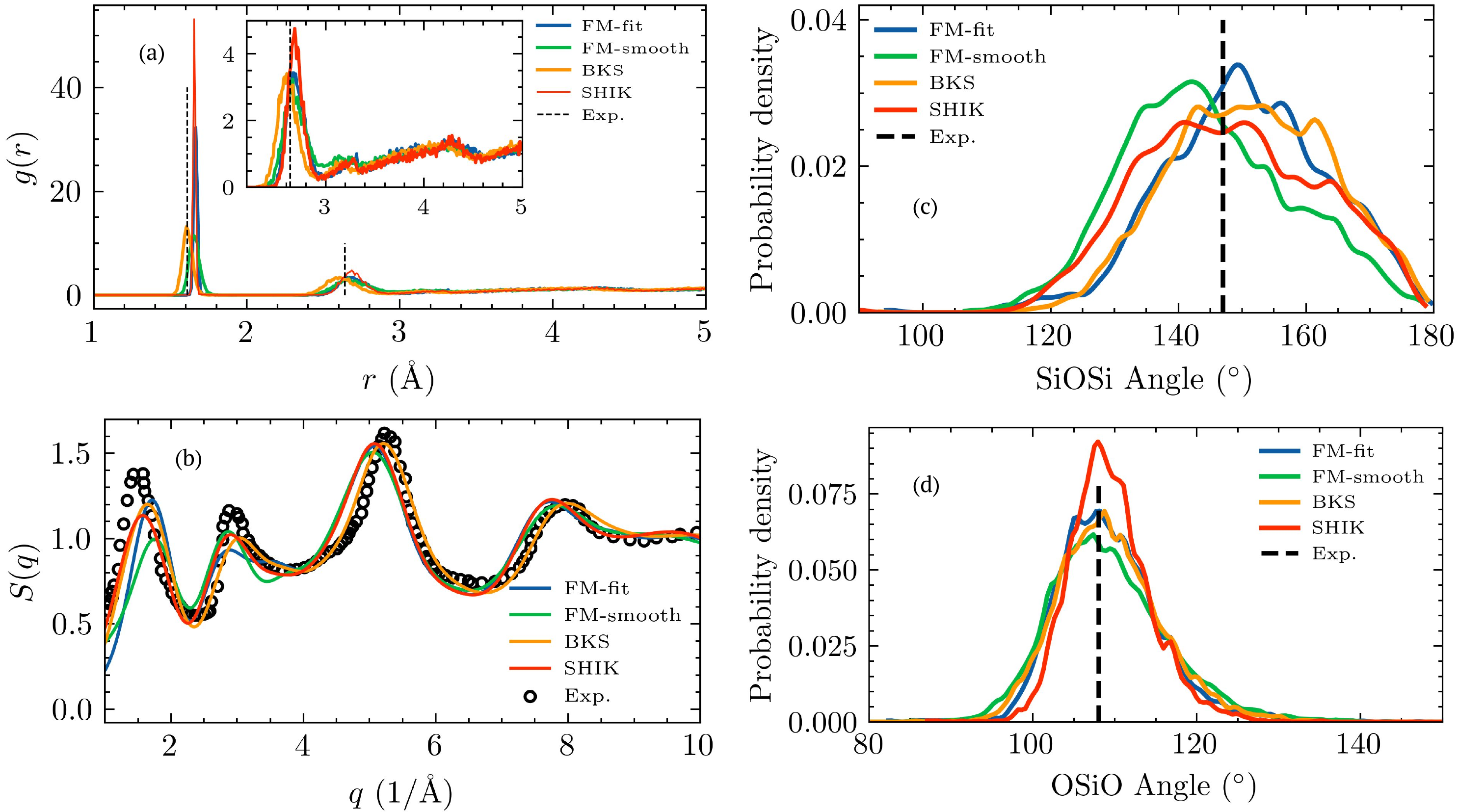}
  \caption{Silica glass structures obtained from melt-quenching simulations with the same cooling rate of 1K/ps by different potentials. In general, the FM potentials reproduce reasonable glass structures similar to other potentials. But without a reliable reference structure at this large cooling rate (the experimental cooling rate is orders smaller), it is difficult to judge which potential is better based on these.}
  \label{fig:s1_lT}
\end{figure}

% \section*{FM-DFT force fields of silicates}

\begin{figure}
  \centering
  \includegraphics[width=1\linewidth]{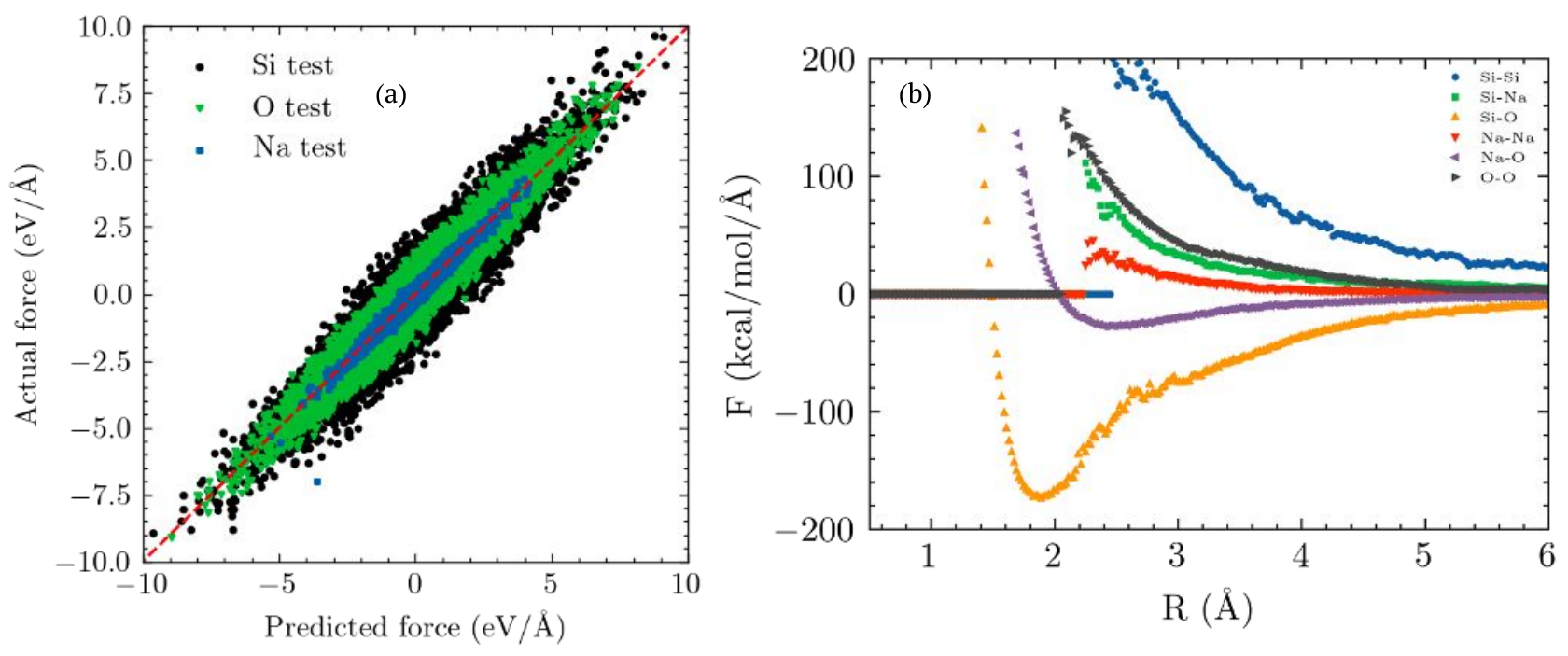}
  \caption{(a) Predicted forces by the ML model vs. the DFT forces for Si, O, and Na atoms from the test dataset. (b) Pair forces as functions of distance in sodium silicate obtained from weights in the ML model.}
  \label{fig:sodium_silicate}
\end{figure}

\begin{figure}
  \centering
  \includegraphics[width=1\linewidth]{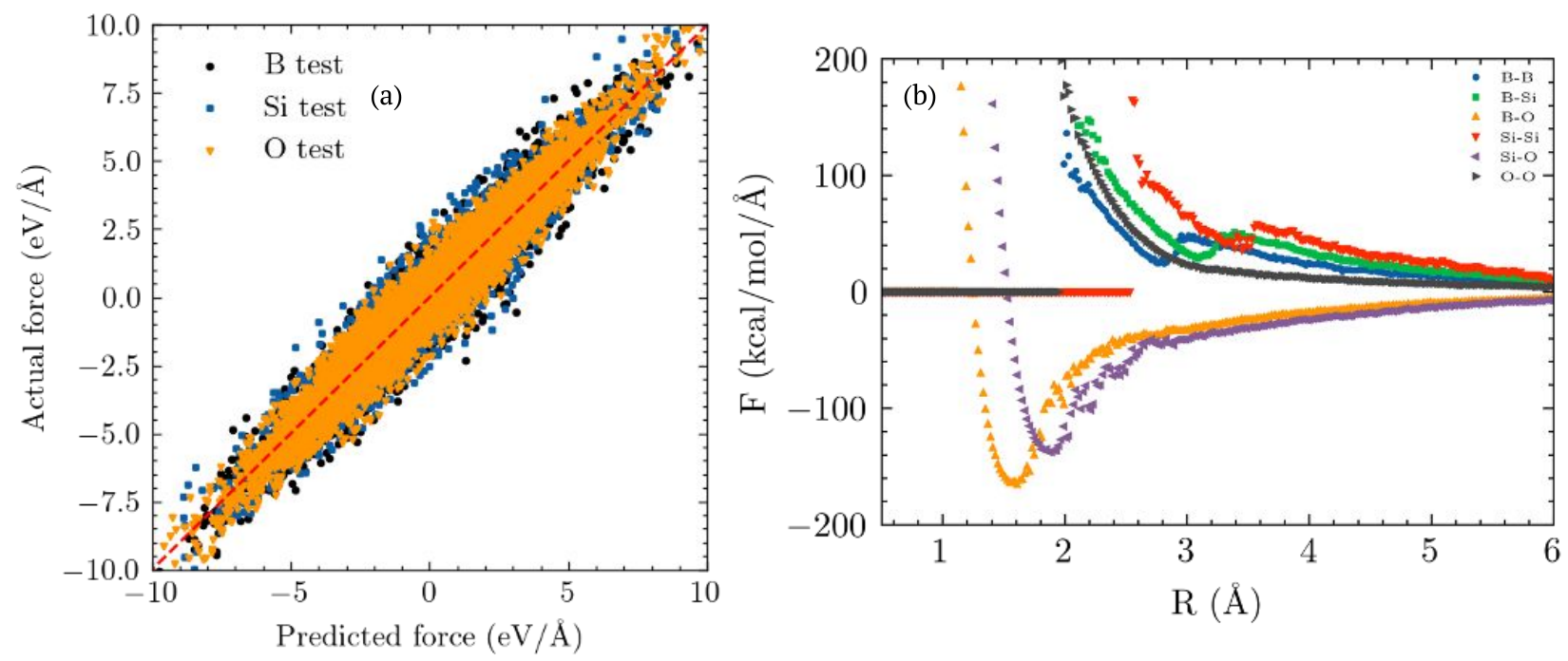}
  \caption{(a) Predicted forces by the ML model vs. the DFT forces for Si, O, and B atoms from the test dataset. and (b) Pair forces as functions of distance in borosilicate obtained from weights in the ML model.}
  \label{fig:borosilicate}
\end{figure}

\bibliography{}